\documentclass{acm}
\usepackage{graphicx}
\usepackage{authblk}
\usepackage{algorithm}
\usepackage{algpseudocode}
\usepackage{listings}
\usepackage{subcaption} 
\usepackage{url}
\usepackage{wrapfig} 
\usepackage{balance}  

\def\embf#1{\textbf{\textit{#1}}}

\begin{document}

\title{DTranx: A SEDA-based Distributed and Transactional Key Value Store with Persistent Memory Log}
\author{Ning Gao}
\author{Zhang Liu}
\author{Dirk Grunwald}
\affil{University of Colorado Boulder\\
	\{Ning.Gao, Zhang.Liu, Dirk.Grunwald\}@colorado.edu }
\date{}
\maketitle

\begin{abstract} 
Current distributed key value stores achieve scalability by trading off
consistency. As persistent memory technologies evolve tremendously,
it is not necessary to sacrifice consistency for performance. 
This paper proposes DTranx, a distributed key value store
based on a persistent memory aware log. DTranx integrates
a state transition based garbage collection mechanism in the log design
to effectively and efficiently reclaim old logs. In addition, DTranx adopts 
the SEDA architecture to exploit higher concurrency in multi-core
environments and employs the optimal core binding strategy to minimize context switch overhead.
Moreover, we customize a hybrid commit protocol that combines optimistic concurrency
control and two-phase commit to reduce critical section of distributed locking
and introduce a locking mechanism to avoid deadlocks and livelocks.\

In our evaluations, DTranx reaches 514.11k 
transactions per second with 36 servers and 95\% read workloads. 
The persistent memory aware log is 30 times faster than 
the SSD based system. And, our state transition
based garbage collection mechanism is efficient and effective.
It does not affect normal transactions and log space usage is
steadily low.
\end{abstract}

\section{Introduction}
There are numerous storage management systems, such as distributed RDBMS,
NoSQL database, distributed file systems, and transactional key value stores.
These systems offer different
levels of transactional and data schema support. Distributed RDBMS
provides strict data schema and full ACID properties with the price of
low availability and efficiency. NoSQL databases and distributed file systems, such
as Cassandra~\cite{cassandra} and GFS~\cite{GFS}, are
scalable and highly available, but they often lack consistency support.
Transactional key value stores, such as BigTable~\cite{BigTable}, sacrifice data schema flexibility, but they offer
higher availability, superior performance, and better scalability.

In this paper we present DTranx, a SEDA-based distributed transactional
key value store with persistent memory log. DTranx follows
the SEDA\cite{seda} architecture to exploit the high concurrency in
multi-core environments. SEDA organizes the software in a network of stages 
where stages contain both the application logic and communication channels.
DTranx adopts lock free queues as the communication channels to reduce contention
among threads. In addition, DTranx binds threads to physical cores to minimize the 
context switch overhead.

Unlike most existing key value stores, DTranx is fully ACID compliant supporting
serializability. To serialize concurrent transactions, we adopt a hybrid of
Optimistic Concurrency Control(OCC) and Two-Phase Commit(2PC) to narrow down
the critical section of distribute locking to the commit time and enables parallel
validation for high scalability.
Furthermore, we avoid deadlocks and livelocks with a customized locking mechanism 
where transactions are aborted if shared lock requests are rejected and
the exclusive lock requests are blocked for a timeout if not granted immediately.
However, if the data is exclusively locked when the new exclusive lock requests come,
the new requests are rejected immediately.

Moreover, DTranx integrates a modular Write-Ahead Log (WAL) which can be configured to use
conventional SATA SSDs or Non-Volatile Memory(NVM)~\cite{Wang:2014:SLT:2732951.2732960} technologies.
Applying NVM in the WAL considerably cuts down the durability cost that most ACID-compliant systems suffer.
A state transition mechanism to garbage collect(GC) WALs is also developed to reclaim the
logs of the completed transactions. The garbage collection process does not affect 
normal transactions since old logs and the current appending log are not in the same file.

In summary, our contributions are as follows:
\begin{itemize}
	\item Adopting SEDA concurrent architecture and employing the optimal core binding strategy;
	\item Customizing a hybrid commit protocol combining optimistic concurrency control and two-phase
	commit and introducing a locking mechanism to avoid deadlocks and livelocks;
	\item Adopting NVM using the Linux {\tt pmem} library in the WAL of the distributed
	transactional system to reduce the persistence overhead and to offer durability;
	\item And, designing a state transition based garbage collection
	mechanism to efficiently reclaim increasing log space without
	affecting normal transactions.
\end{itemize}

\section{Background}
\begin{figure}
	\begin{center} 
		\includegraphics[trim = 0mm 0mm 0mm 0mm, clip=true, scale=0.42]{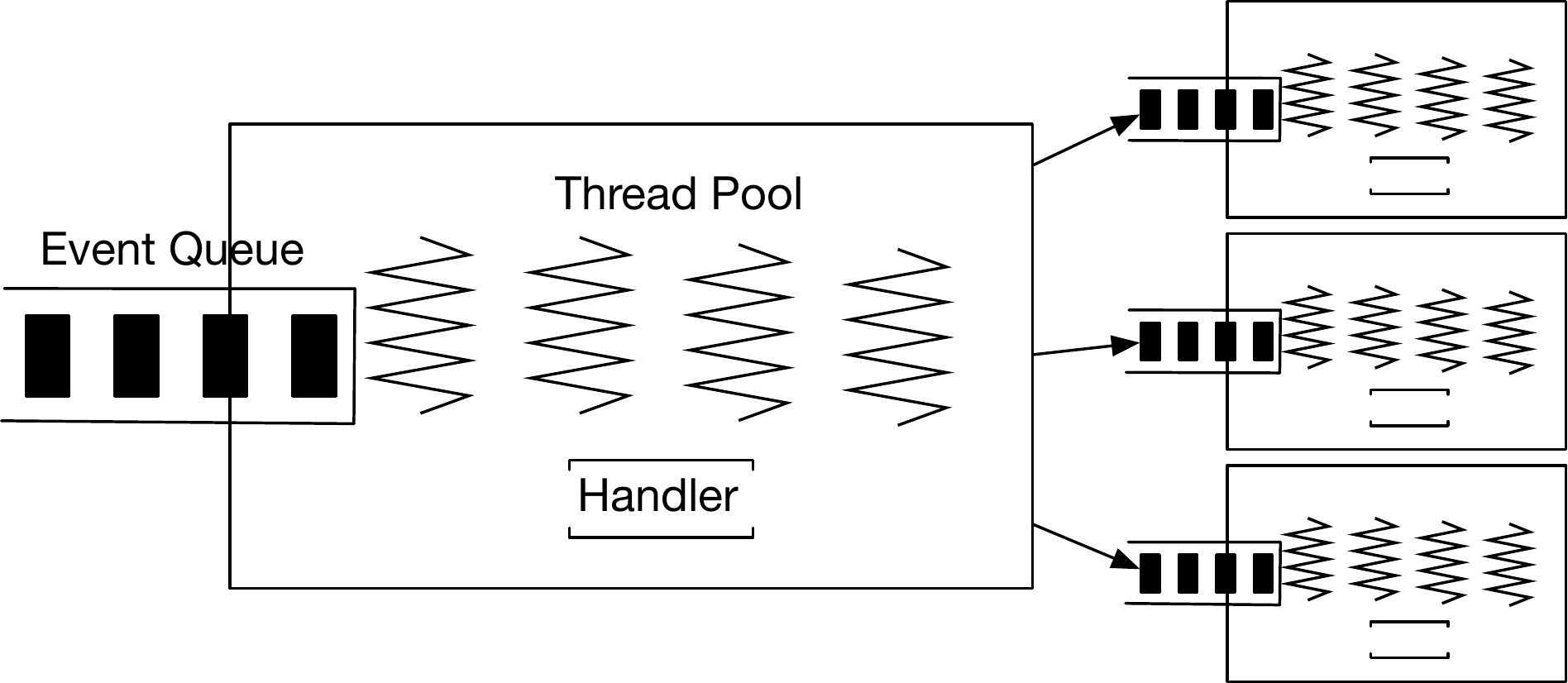} 
		\caption{SEDA architecture}
		\label{fig:seda-basic} 
	\end{center}
	\vspace{-5mm} 
\end{figure}
\textbf{Staged Event-Driven Architecture} SEDA is a highly concurrent architecture, consisting of
a network of event-driven stages connected by queues. A stage
is an independent software module that manages a shared resource. 
For example, the lock service in transactional systems is a stage that maintains the
locking information and handles lock requests. As shown in Figure\ref{fig:seda-basic},
a stage is composed of an incoming event queue, a handler, and 
a thread pool. Besides the three core elements, 
SEDA adds a controller to adjust the thread pool size dynamically.
The event handler sends events to another stage by invoking the enqueue
operation on the incoming event queue of that stage. SEDA
brings four benefits. First, it offers modularity and independent load management. 
Second, it facilitates debugging and performance analysis, which has
always been a tough task for multi-threaded programs.
Third, it optimizes the overall system performance by dynamically adjusting
resource allocations, such as thread numbers among stages. Fourth, it
enables batch request processing. For example, a database stage could
write multiple keys at a time. However, SEDA requires nonblocking design of the event handler.

\textbf{Optimistic Concurrency Control}
Concurrent control is the coordination of multiple concurrent accesses
to the database and Philip et. al~\cite{concurrency_control} decomposed
it into two major subproblems: read-write synchronization and write-write synchronization.
There are pessimistic and optimistic approaches towards both subproblems.
The pessimistic approach assumes the probability of access conflicts to be
high and decides whether to restart at the start of transactions, such as two-phase locking. 
The optimistic approach assumes the probability
of access conflicts to be low and decides whether to restart at the end of transactions.
Specifically, Optimistic Concurrency Control(OCC)~\cite{Harder:1984} consists of three phases:
read, validation, and write. During the read phase, transactions read databases and
store updated data in the buffer. Then, databases check whether the current transaction
is in conflict with any concurrent operations. Finally, if it passes the validation
phase, the current transaction proceeds to update the database states.   

\textbf{Two-Phase Commit}
Two-Phase Commit(2PC) is a classic commit protocol in the distributed environment that 
guarantees agreement among servers on the commit results. Moreover, once the agreement
is reached, the commit results hold however the servers fail. There are two roles for the
servers: coordinator and participant. During the first phase, coordinators initiate 
2PC by sending \embf{prepare} messages to participants and participants either accept
or reject the \embf{prepare} messages. In the second phase, coordinators send out
\embf{commit} messages if all participants accept the \embf{prepare} messages and 
\embf{abort} messages, otherwise. Both coordinators and participants write Write-Ahead
Log(WAL) to persist volatile states, such that the commit decisions for recovery purposes.


\section{Design}
\label{sec:design}
DTranx adopts the SEDA architecture to reach the optimal performance in each server
and achieves serializability by combining OCC and 2PC protocols. Furthermore,
it introduces a NVM-based WAL design with a garbage collection mechanism to effectively and
efficiently reclaim logs.
\subsection{Architecture Overview}

\begin{figure}
	\begin{center} 
		\includegraphics[trim = 55mm 45mm 0mm 29mm, clip=true, scale=0.4]{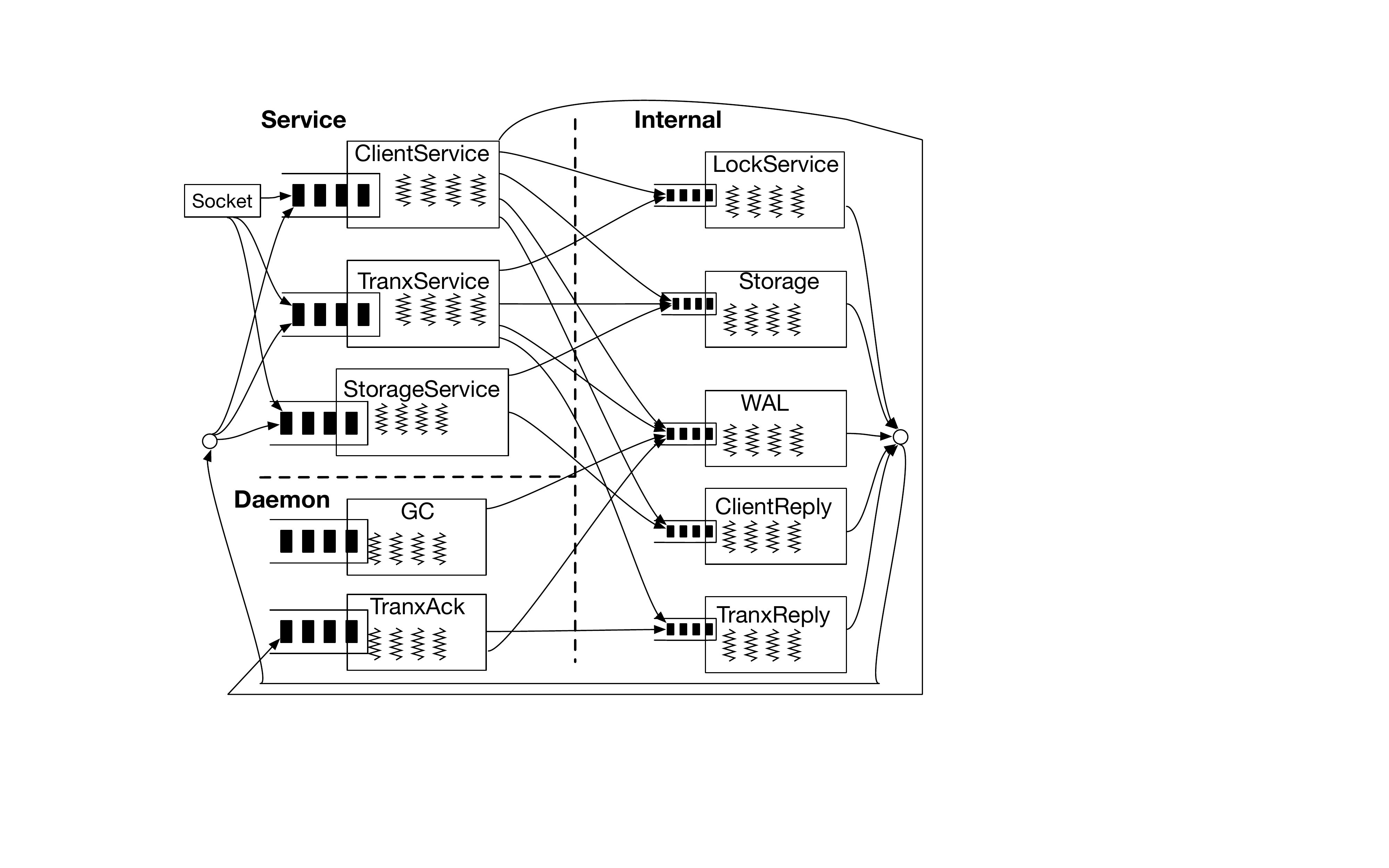} 
		\caption{DTranx architecture.}
		\label{fig:seda} 
	\end{center} 
	\vspace{-5mm}
\end{figure}
As shown in Figure\ref{fig:seda}, DTranx follows the SEDA~\cite{seda} design and invents
three categories of stages: \embf{Service}, \embf{Internal},and \embf{Daemon}. \embf{Service}
stages handle Remote Procedure Calls(RPCs). For example, \textit{ClientService} accepts transaction
commit requests from clients and \textit{TranxService} processes 2PC requests from peer servers.
\embf{Internal} stages manage local shared resources. For example, \textit{LockService} maintains
locking states and \textit{WAL} writes logs to persistent storage. \embf{Daemon} stages
run background tasks. For example, \textit{GC} periodically reclaims logs and \textit{TranxAck}
sends commit results from coordinators to participants. 

To further exploit concurrency in SEDA, DTranx adjusts each of the stage components. 
First, DTranx removes the dynamic control of the thread pools but statically assigns 
thread numbers for each stage, after which threads are bound to physical CPU cores. 
We found out that one thread for each stage yields better performance when context switching
is rare than that of multiple threads for each stage when context switching happens frequently.
However, dynamic control of the thread pools is enabled in certain stages where handlers might be
blocking. For example, \textit{Storage} launches multiple threads to handle I/O requests which
involve blocking system calls. Besides the core bindings for DTranx threads, 
kernel Interrupt Request(IRQ) threads are bound to CPU cores as well since I/O throughput
is severely affected otherwise.

Second, DTranx adopts lock free queues as the incoming event queues such that the enqueue/dequeue
operations on the queues are nonblocking and it achieves high throughput without compromising consistency.
The lock free queues utilize atomic primitives to reserve a spot and then proceed to read/write in 
non-critical sections. In addition, multiple queues are created in each lock free queue to spread loads.

Third, DTranx reduces queue element construction and destruction costs by pushing element pointers
, instead of the element itself into the lock free queues and allocating an element pool to store destructed
elements. For example, \embf{Service} stages get elements from the pool when new requests come and
\embf{Internal} stages put elements to the pool when requests are completed.

\subsection{Serializability}
\begin{figure}
	\begin{center} 
		\includegraphics[trim = 0mm 0mm 0mm 0mm, clip=true, scale=0.38]{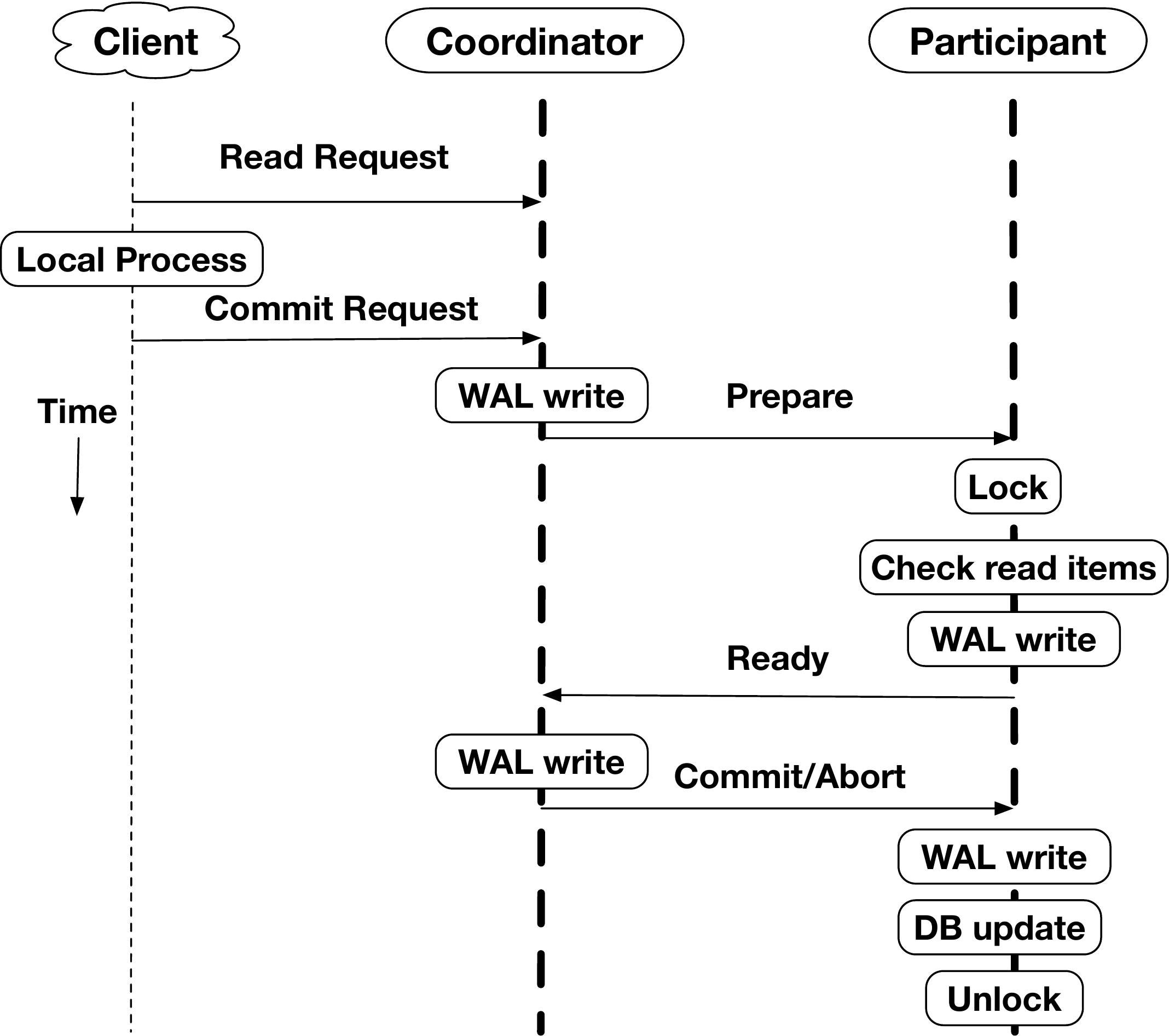} 
		\vspace{+2mm}
		\caption{Commit Protocol.}
		\label{fig:commit_protocol} 
	\end{center} 
	\vspace{-8mm}
\end{figure}
DTranx combines OCC and 2PC protocols to guarantee serializability, following
Alexander's~\cite{docc2} hybrid OCC scheme that embedded lock acquisition
and validation in the 2PC. The main benefit compared to distributed Two-Phase 
Locking(2PL) is that locks are only held during the commit time and DTranx
employs parallel validation for better scalability. The detailed protocol
flows are shown in Figure\ref{fig:commit_protocol}. Additionally, if all data items in a transaction
are stored in the same server, 2PC is converted to One-Phase Commit(1PC) to reduce latencies.

Initially (Stage 1), transactions read data without locks
and clients keep track of the read items, read item versions, and write items in the local buffer.
At commit time (Stage 2), clients choose a server
as the coordinator and send it the transactions. 
During the first phase, coordinators initiate
2PC by first sending prepare messages to participants. 
Then, participants lock both the read
and write items, check the read item versions, write WAL logs
and reply to the coordinator. During the second phase, coordinators wait for
responses from all participants, then decides whether to commit or abort, 
and notifies all participants of the agreed result. However, if any participant aborts in the first phase,
the coordinator immediately sends out abort messages without waiting for all replies. Finally,
participants write WAL logs, update database states, and unlock all relevant data.

\textbf{Proof of Serializability}

\textbf{\textit{Assumption}}: Two phase locking(2PL) ensures serializability,
see proof at \cite{dbcomplete}.

\textbf{\textit{Method}}: We reduce the hybrid OCC to 2PL.  Using action
abbreviations L~(Locking), C~(Checking), U~(Unlock), R~(Read), W~(Write) and object abbreviations r~(read items),
w~(write items). Concatenated action and object symbols represent tasks, e.g., Lr means
``lock read items''. The sequencing abbreviation ``-'' binds two actions and enforces an ``execute
before'' local order and $\rightarrow$ binds two tasks and enforces an
``execute before'' distributed order. Our transactions can thus be
represented as R $\rightarrow$ Lrw-Cr $\rightarrow$ W-Urw. The Cr action validates the
read items. If any read item has been changed after it was read, the transaction aborts,
releasing all locks. If not, our successful transaction is equivalent to Lr-R
$\rightarrow$ Lw-Cr $\rightarrow$ W-Urw, thus Lr-R $\rightarrow$ Lw
$\rightarrow$ W-Urw. In this way, all locking actions precede all unlocking
actions, which is 2PL. Unlocking after committing to the
database avoids cascading rollbacks. For successful transactions, the
serialization point is the moment when all the write locks are granted.

\textbf{Deadlock}
Common deadlock avoidance methods are timeout, wait-for
graph, ordered locking and timestamps with wait-die or
wait-wound mechanisms. SiloR~\cite{silor} avoids deadlocks by enforcing
a global order on the locking sequences, necessitating multiple round trips in distributed environments.
The wait-for graph introduces too much network traffic and
timestamps method requires a global synchronized clock, which 
will become the bottleneck or single point
of failure. Our deadlock avoidance method aborts transactions immediately 
if read locks are not granted and waits for a configurable
time period(e.g. 50ms) before aborting write lock requests. However, if
the data is currently exclusively locked, the write lock request is 
aborted immediately. If a transaction is aborted since write
locks are not granted, DTranx retries committing it after an exponential timeout. 
If a transaction is aborted since read locks are not granted, DTranx
restarts it immediately. This is because read lock request denial indicates there
are concurrent transactions updating the same item and retrying
committing will fail again. 

\textbf{Livelock} Write starvation rarely happens since write lock requests are blocked
for a short fixed period and exponential backoff technique is adopted to reduce the probability of
lock conflicts when the transactions are retried.
The same goes for read starvation since the number of read items are usually much larger
than write items. 

\subsection{Persistent Memory}
\begin{figure*}[bth]
	\begin{center}
		\begin{tabular}{ccc}
			\begin{minipage}{3.3in}
				\begin{center} 
					\includegraphics[trim = 55mm 0mm 0mm 8mm, clip=true, scale=0.35]{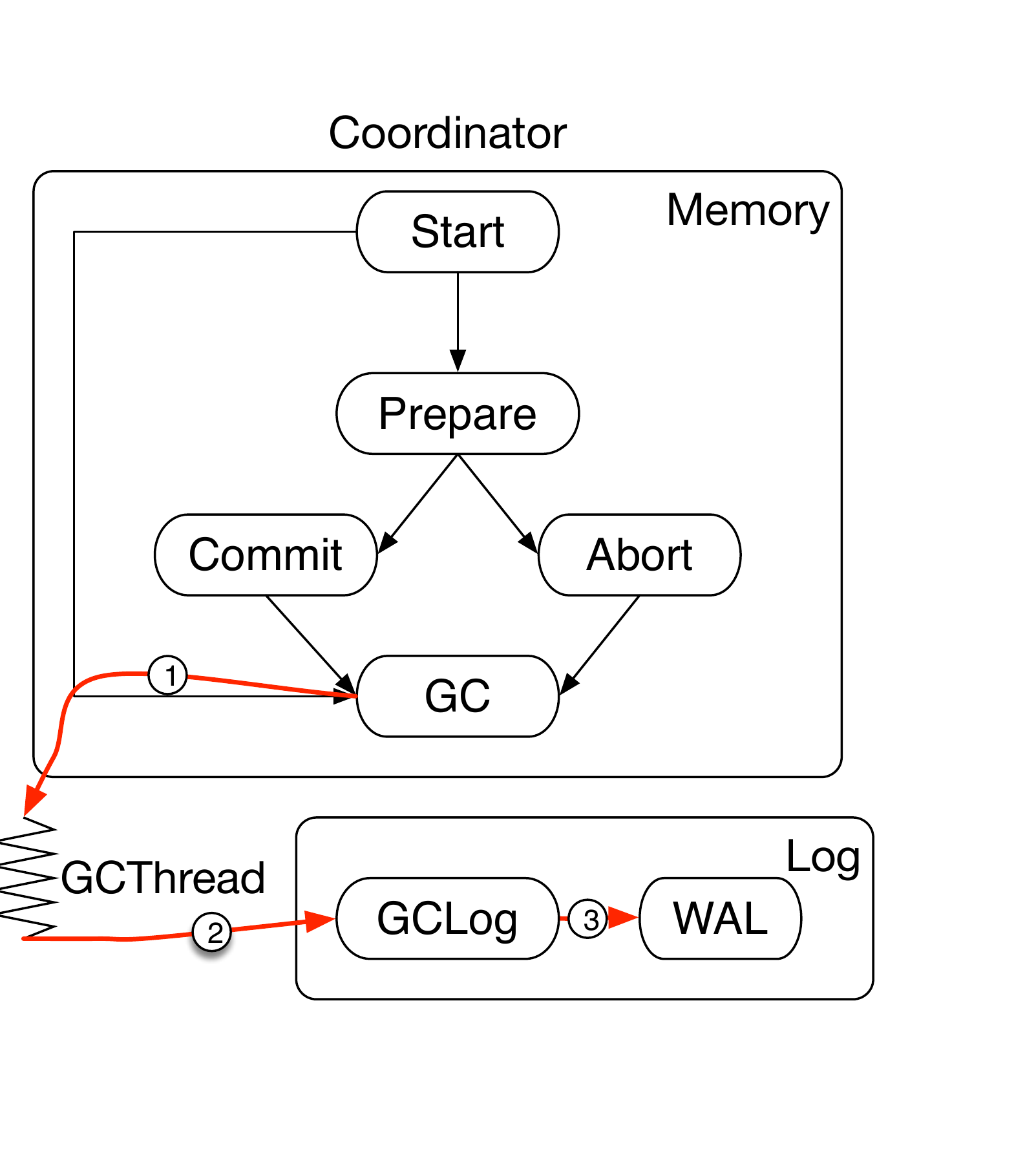} 
				\end{center}
			\end{minipage}
			&&
			\begin{minipage}{3.3in}
				\begin{center} 
					\includegraphics[trim = 35mm 0mm 0mm 5mm,
					clip=true, scale=0.35]{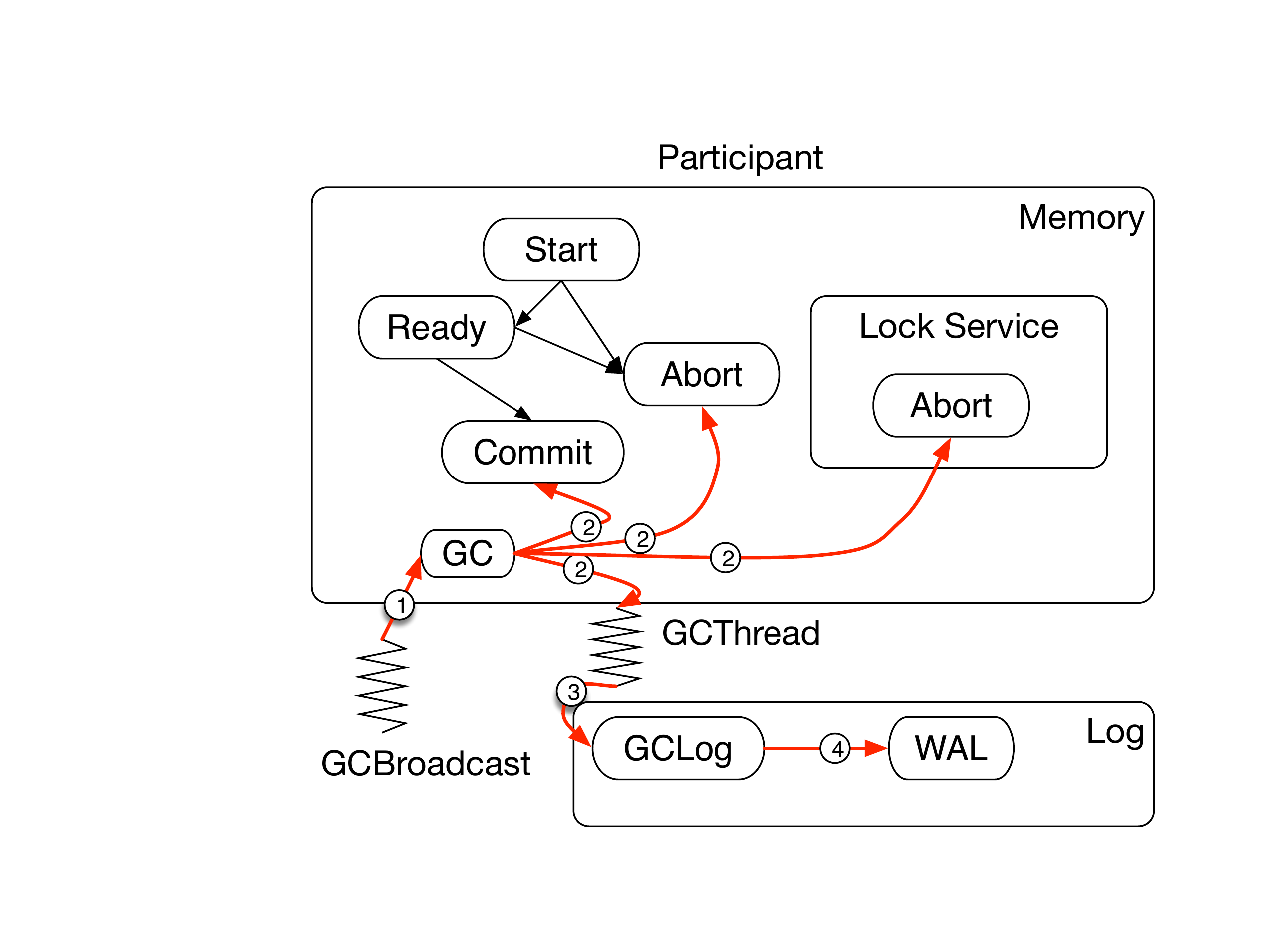} 
					
				\end{center} 
			\end{minipage}
		\end{tabular}
		\caption{Coordinator and Participant state transition. Black arrows show the state transitions 
			and red arrows show the ordered steps of garbage collection.}
		\label{fig:state_transition} 
	\end{center}
	\vspace{-5mm}
\end{figure*}

In distributed systems, the logging module plays a critical role in failure recovery. 
WAL is the persistent copy of the volatile states that are subject to failures due to
power outage and kernel hanging. However, persisting WAL to the durable storage results in
long latencies. With the advances of the NVM technologies, the performance gap between in-memory
and persistent storage accesses is narrowing. Thus, we propose a WAL design based on NVM
and introduce a garbage collection mechanism to effectively and efficiently reclaim the 
limited NVM space.

\subsubsection{Log Design}
The logging module is designed in three vertical layers: NVM library, LogManager 
and TranxLog. The NVM library provides the basic interface to persistent memory
to create files, read and write data. LogManager structures the log into a list of
log files, calculates checksums, and supports block read/write operations.
Lastly, TranxLog offers high level abstractions for distributed transaction logs and 
presents a continuous and append-only log.

We use Intel's NVM~\cite{pmem} library to manipulate memory 
mapping for log files in persistent memory. After log files are mapped
to the memory space, writes are immediately durable after being flushed from
cache to memory(e.g. using clflush in the x86 instruction set).
Two adjustments of the NVM library are made. First, a read pointer is added to 
the original NVM library to provide an on-demand read interface. 
Second, internal write locks are disabled since only one thread is launched in the
\textit{WAL} stage, thus no race conditions.

On top of NVM library, LogManager organizes the logs in a list structure such that logs of
variable sizes are supported. To reduce I/O system calls and reach higher
throughput, reads and writes are block based and logs in the same block are buffered
in memory. In addition, checksums are calculated and written for each block to detect
data corruption. 

TranxLog serializes transaction logs such as \textit{CommitLog} that
records commit states for coordinators and \textit{ReadyLog} that records
ready states for participants. Then, TranxLog separates WALs into files
whose names are set to their creation timestamps. Thus, the file with the smallest
timestamp is the oldest one, with which the garbage collector starts. On the other hand, reclaiming
old log files does not interfere with current transactions since current transactions are 
appending logs to new log files.

\subsubsection{Garbage Collection}\label{subsubsec:garbagecollection}
Since WAL is written as transactions are committed, its size would increase indefinitely if DTranx
does not reclaim the WAL of complete transactions.
WAL for transactions that reach consensus are not required during recovery.
Therefore, we introduce a state transition based garbage collection mechanism to identify 
unnecessary logs without performance hiccups. 
In particular, each transaction is assigned with a unique TranxID that combines 
the ServerID and a local monotonically increasing 64-bit integer.
Since ServerIDs are assigned as the server indexes in the group 
membership stored in the replicated state machine Raft~\cite{raft}, TranxIDs are guaranteed 
to be distinct among servers.
Moreover, each server keeps updating the largest committed TranxID, LC\_TranxID, where
transactions with TranxIDs less than LC\_TranxID have reached consensus.
Then, each server broadcasts its LC\_TranxID and stores the LC\_TranxIDs from other servers
in a fixed-size GC log.  The benefits of the GC log are twofold: 
it is fixed size space usage and it enables WAL reclamation. 

The state transition flow is illustrated in Figure \ref{fig:state_transition}.
On the one hand, each transaction has a state to represent the current stages in the 2PC and 
each server has a \textit{GC} state to record completed transactions where LC\_TranxIDs are calculated. 
On the other hand,  
there are volatile and nonvolatile states where the nonvolatile
states are durable copies of the volatile ones. 
For example, WALs are the nonvolatile copies of 2PC states including
\textit{Start}, \textit{Prepare}, \textit{Ready}, \textit{Commit}, and \textit{Abort}.
GCLog is the nonvolatile state persisting the \textit{GC} state.
Although both WALs and GCLog are persistent copies of the volatile states, their orders of updating volatile and nonvolatile
states differ. For WALs, nonvolatile 2PC states are updated after WALs are written in order for the transactions to be recoverable.
For the GCLog, \textit{GC} state is updated before GCLog for two reasons. 
First, the history can be replayed as long as WALs are not reclaimed yet. Second, 
accumulating in-memory \textit{GC} states and writing to the GCLog in batch is more I/O efficient.
For coordinators, GCThread periodically collects the volatile \textit{GC} states and updates the GCLog, after which
WALs containing only completed transactions are reclaimed and the LC\_TranxIDs are broadcasted to all
the other servers. For the participants, GCBroadcast thread passively receives
the broadcasted LC\_TranxID, updates the local \textit{GC} state and GCLog,
and then reclaims WALs. 

Not only does the state transition help to reclaim WALs, it is also utilized to clean the aborted
transaction IDs in the lock service, which are referenced to avoid faulty re-lock situations. For example, after
a participant receives the prepare request of transactionA and its volatile state is checked to be \textit{Start}, 
an abort request of transactionA arrives, changing the volatile state to \textit{Abort}.
It is possible that abort requests come before prepare requests are done since the coordinator
immediately sends out abort requests if any participants aborts. 
Note that these two requests are processed concurrently.
Then, the prepare requests lock the data items and these locks will never be unlocked.
Nonetheless, the committed transaction IDs are not stored in the lock service since coordinators only
send out commit requests after all participants agree to commit, in which case it is impossible that
prepare and commit requests are processed concurrently.

\section{Implementation}
\label{sec:implementation}
In this section, we explain the implementation details that optimize
DTranx performance.
\subsection{Cache}
DTranx enables client side cache to avoid excessive network traffic.
The caching policy works as follows: (1) Data cache is updated if read or commit requests succeed. 
(2) Data cache is invalidated if commit requests fail. In addition, DTranx servers piggyback the 
updated data in the response to failed commit requests such that the clients can update the local
cache and read requests in the retrying transaction can read from the local cache.

On the other hand, DTranx enables server side database cache to serve read
requests in lower latency and it adopts the write-through strategy for durability.

\subsection{Exactly-Once RPC}
There are three different RPC calls corresponding
to the three stages in Figure~\ref{fig:seda}:
Read requests from clients to servers; Commit requests from
clients to servers; and, Transaction requests from servers
to servers. 
Duplicate processing of RPCs would lead to system failures in certain cases. For example,
if DTranx servers process a duplicate prepare request after the corresponding 
commit request is done, the locking service would lock the data items and future transactions
would not be able to update these data. Therefore, DTranx should guarantee
to process each RPC exactly once.
 
First, we guarantee at least once delivery by resending messages on the sender side if
no responses are received within a timeout. We build the RPC protocol based 
on the ZeroMQ library, which automatically resends messages if they are lost.
In addition, DTranx implements the retrying mechanism itself when no responses are received since 
at least once delivery in ZeroMQ does not indicate at least once delivery in DTranx. 
For example, if servers are restarted after the ZeroMQ library receives a message but before the DTranx system
detects the message, ZeroMQ does not retry the message and the message is lost.

Second, we guarantee at most once processing by blocking duplicate messages. 
we assign distinct IDs for each RPC message and receivers record
the IDs of completed messages. Read requests are never blocked since they are idempotent. For Commit requests, each
message has a clientID and messageID where the clientID is distinct
for each TCP connection and the messageID is monotonically increasing
for each client. ClientIDs are assigned by the ZeroMQ library when the connection
is established. For Transaction requests, each message has a unique transaction ID(TranxID) and 
a message type. TranxID is the concatenation of
the distinct coordinator server ID and a monotonically increasing integer.
And, there are four message types corresponding to the four Transaction requests in Figure~\ref{fig:commit_protocol}:
\textit{Prepare}, \textit{Ready}, \textit{Commit}, and \textit{Abort}.
With at least once delivery and at most once processing, each message is processed exactly once. 

\subsection{LevelDB}
We choose levelDB\cite{leveldb} as the local database implementation
since it is lightweight and efficient compared to multi-version KV stores.
To validate the read items during the OCC commit, DTranx keeps a version number
for each key value pair by storing the combination of the real value and
a version number as the value in levelDB. 
The real value and the version number are separated by a special delimiter, such as
\textit{\#}. When clients send read requests, servers interpret the values retrieved from
levelDB and returns the value and the version number to the clients.  
When clients send Commit requests, servers increment the corresponding 
version numbers by 1 if transactions commit.

\subsection{Fault Recovery}
As the cluster size increases, the probability of
server failures will increase considerably. For example, if the aggregated
MTBF~(Mean Time Between Failure) of a server is 1 year including disk
failures, network failures etc., then in a cluster of 100 servers, there
is a server failure every 3 to 4 days on average.
DTranx triggers the recovery process in two stages: local recovery and global recovery. 
Local recovery reapplies local logs by updating databases if transactions commit and
lock data items if no agreement has been reached. 
In addition, DTranx fills the TranxID space with aborted transactions. 
Missing transactions are possible when servers 
crash immediately after read item checking fails in coordinators. 
Global recovery repairs transactions of which commit results can not be decided unilaterally.
It is initiated after local recovery to inquire transaction states from other involved servers.
Specially, if the coordinator is in \textit{Prepare} state and all participants are in \textit{Ready} states, 
neither committing nor aborting violates distributed consensus. DTranx
chooses to abort them such that the clients can assume the transaction failure
if no responses are received. 

On the other hand, DTranx starts service stages in Figure~\ref{fig:seda} after local recovery
such that the changes from completed transactions are applied and in-memory states of
ongoing transactions are stored. However, the order between service stage startup and
global recovery does not matter and DTranx chooses to starts service stages before global
recovery to reduce the service downtime.

\subsection{Optimization}
In order to achieve better performance, multiple optimization techniques are applied.
The most significant techniques are listed below. 

\begin{itemize}

\item \textbf{Delayed In-Memory Reclamation} DTranx reclaims the volatile \textit{Commit}/\textit{Abort} 
states in participants when the servers are under light loads to avoid performance hiccups.

\item \textbf{Batch Ack Phase} DTranx delays the second phase(Ack phase) of 2PC when coordinators
send transaction commit results. We delegate the Ack phase to a separate stage, TranxAck in Figure~\ref{fig:seda},
to reduce the transaction latency and offload the high processing demand of the ClientService stage.

\item \textbf{Core Bindings} We manually analyze the queue size for each stage and bind the 
threads to physical cores in an optimal way. 
The best core binding strategy yields almost 6 times higher throughput than the worst. 
In the future, we plan to explore how to automate the core bindings to attain
the best performance based on the number of CPU cores available.

\end{itemize}

\section{Evaluation}
\label{sec:evaluation}
Our benchmark tests are run on a Cloudlab \cite{cloudlab} cluster
with 36 machines. Each of the machines has Intel E5-2660 v3, 20
2.6GHz cores, equipped with 130GB RAM, 480GB solid state disk at 6GB/sec, and
10Gbps Ethernet card. We emulate the NVM by enabling DAX support in Linux
to create a PM-aware environment. The DRAM based emulation is adopted since current
persistent memory latency is comparable to DRAM and NVM was not available. For example, 
STT-RAM \cite{sttmram} achieves $\sim$10ns write latency compared to 50ns DRAM latency.
NVM throughput is also far beyond the current usage as shown in Figure \ref{fig:pmemspace}.
To generate workloads, we use Yahoo Cloud Serving
Benchmark(YCSB) \cite{ycsb} C++ version
and add DTranx and HyperDex support. YCSB clients
are running in separate servers from the cloud servers that
accommodate the DTranx system. For each test result, the average
of 3 runs are reported. 

\subsection{Environment Setup}
First, we evaluate DTranx with a database of 120 million key value items in a 36-node cluster. 
Test data keys are generated as integers from 1 to 120 million 
and values are 100 bytes of random characters. 
Transactions are categorized into read and update transactions.
Read transactions only contain read items and update transactions contain 1 write item. 
The total number of read/write items in one single transaction is uniformly distributed between 1 and 3.

\subsection{Transaction}
\begin{figure}[t]
	\begin{center} 
		\includegraphics[trim = 0mm 0mm 10mm 4mm, clip=true, scale=0.46]{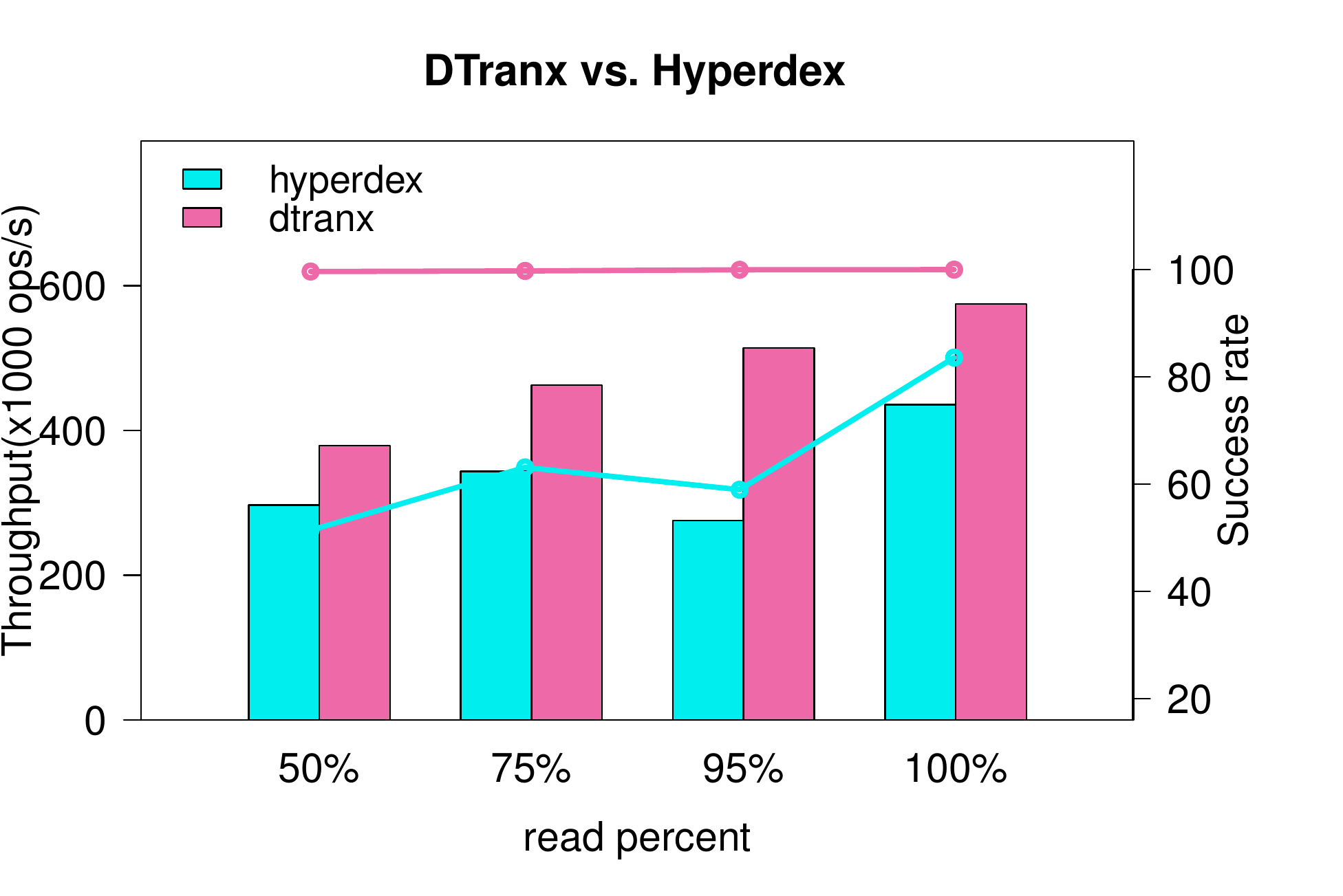} 
		\caption{Distributed transactions. The horizontal axis represents the
			percentage of read transactions while the vertical axis shows the
			throughput on the left and commit
			success rates on the right.}
		\label{fig:hyperdex} 
	\end{center} 
	\vspace{-5mm}
\end{figure}

In Figure \ref{fig:hyperdex}, DTranx is compared with Hyperdex Warp~\cite{warp} that supports 
distributed transactions. Only successful transactions are counted in the throughput metric. 
DTranx shows approximately 30\% higher throughput than Hyperdex and DTranx degrades slowly 
as the percentage of update transactions increases. Moreover, DTranx maintains high commit success rates.
For example, DTranx reaches 
99.65\% success rate for 50\% read
workloads. 
On the other hand, Hyperdex shows high throughput but the software is
unstable and periodically fails due to internal assertion errors, leading to 
low success rates. For example, several servers crashed during the 95\% read workload,
causing 58.96\% success rates and 275.72k ops/sec throughput. To remedy the crash effect,
we restarted the servers manually after each run.
There are three reasons why DTranx outperforms Hyperdex. First, DTranx 
follows the highly concurrent SEDA architecture with lock free queues and stages are bound to physical
cores, utilizing all CPU power and avoiding context switching overhead. Second, DTranx
integrates the NVM based log that bypasses system calls like sync/fsync, reducing 
log persistence latencies. Third, DTranx applies various optimization techniques, such as
an allocated element pool, batch ack phase, and optimal core binding strategy. 
Furthermore, strace~\cite{strace} reveals that Hyperdex does not synchronize data to physical
storage devices immediately after write log calls. 
While the Hyperdex paper supports fault recovery by replication, that version of the software is not 
publicly available. 
Lastly, the average latency for DTranx is
below 2ms when the throughput is 50\% of the maximum and it increases to 10ms 
when the throughput reaches the maximum.

\begin{figure}
	\begin{center} 
		\includegraphics[trim = 0mm 0mm 10mm 7mm, clip=true, scale=0.48]{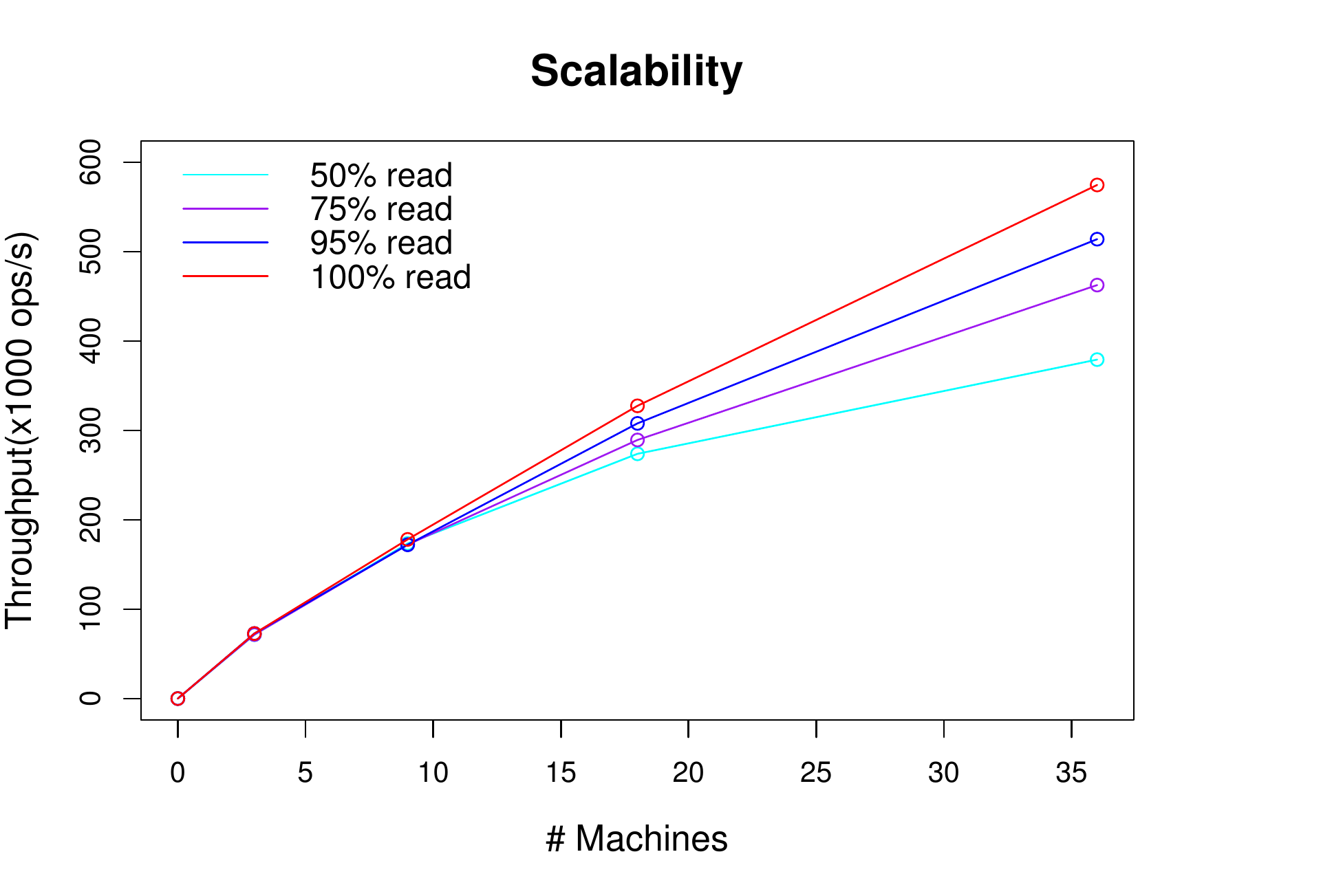} 
		\caption{DTranx Scalability. Four workloads are run, namely 50\%, 75\%, 95\%, and 100\% read workloads.}
		\vspace{-5mm}
		\label{fig:scalability} 
	\end{center} 
\end{figure}
\subsection{Scalability}
In this experiment, scalability tests are run against cluster of 3, 9,
18 and 36 servers. Corresponding to the cluster size, 10, 30, 60, 120 million keys are
inserted into DTranx. As shown in Figure \ref{fig:scalability}, the throughput shows linear
increases as more nodes are involved. For example, with pure read workloads, 
throughput reaches 574.76k reqs/sec with 36 nodes. 
In addition, workloads with various mixture of read and update transactions are benchmarked. 
Even with 50\% read workloads and 50\% update workloads, the throughput is 60\% to 85\% of
that with pure read workloads. The high scalability of DTranx results from our efficient hybrid 
commit protocol design that minimizes the critical section of distributed locking and reduces the 2PC
to 1PC whenever possible.

\subsection{Persistent Memory}
\begin{figure}
		\begin{center} 
				\begin{subfigure}{0.23\textwidth}
						\includegraphics[trim = 0mm 0mm 0mm 0mm, scale=0.38]{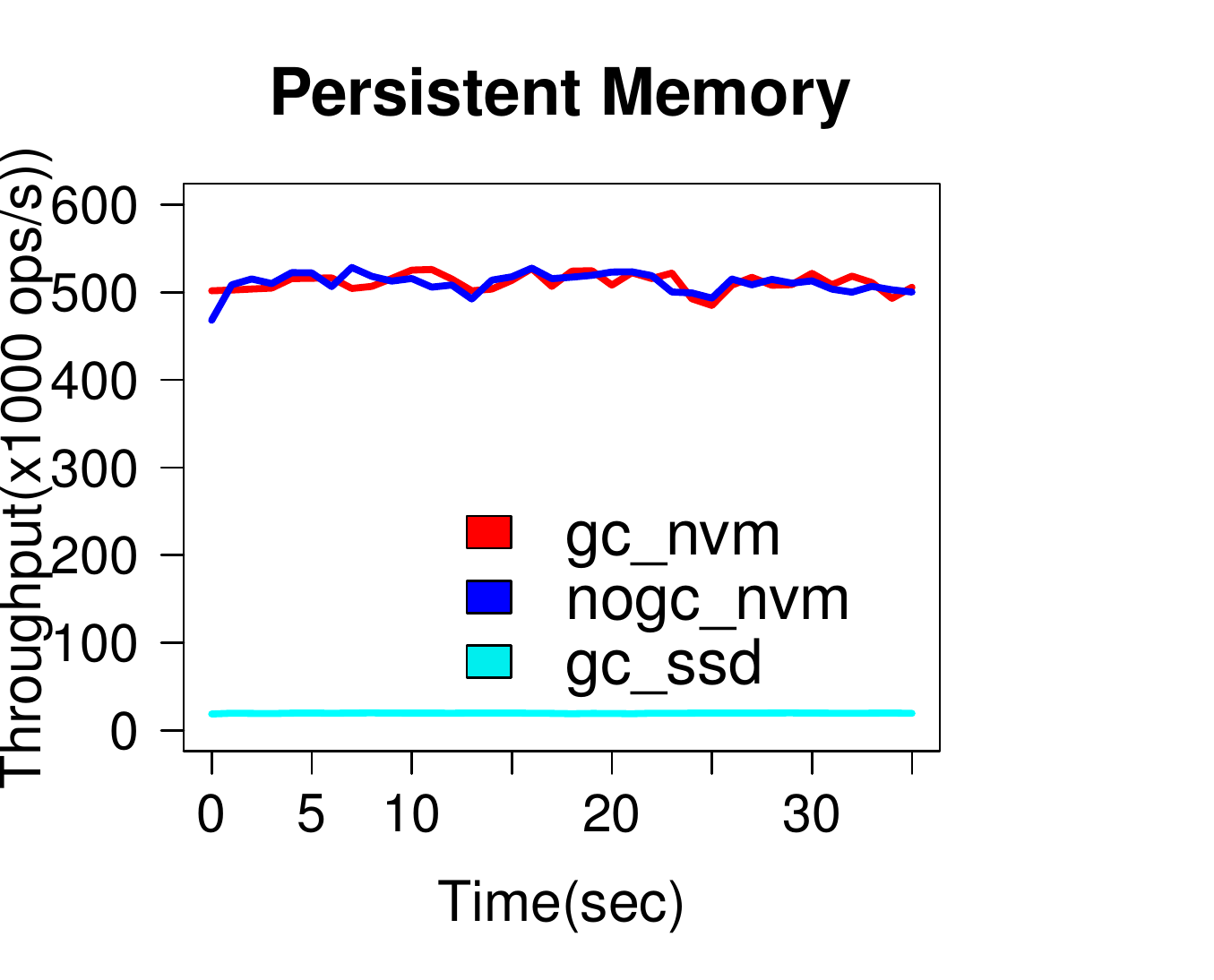}
						\caption{throughput}
						\label{fig:pmemthroughput}
					\end{subfigure}
				\begin{subfigure}{0.23\textwidth}
						\includegraphics[trim = 0mm 0mm 0mm 0mm, scale=0.38]{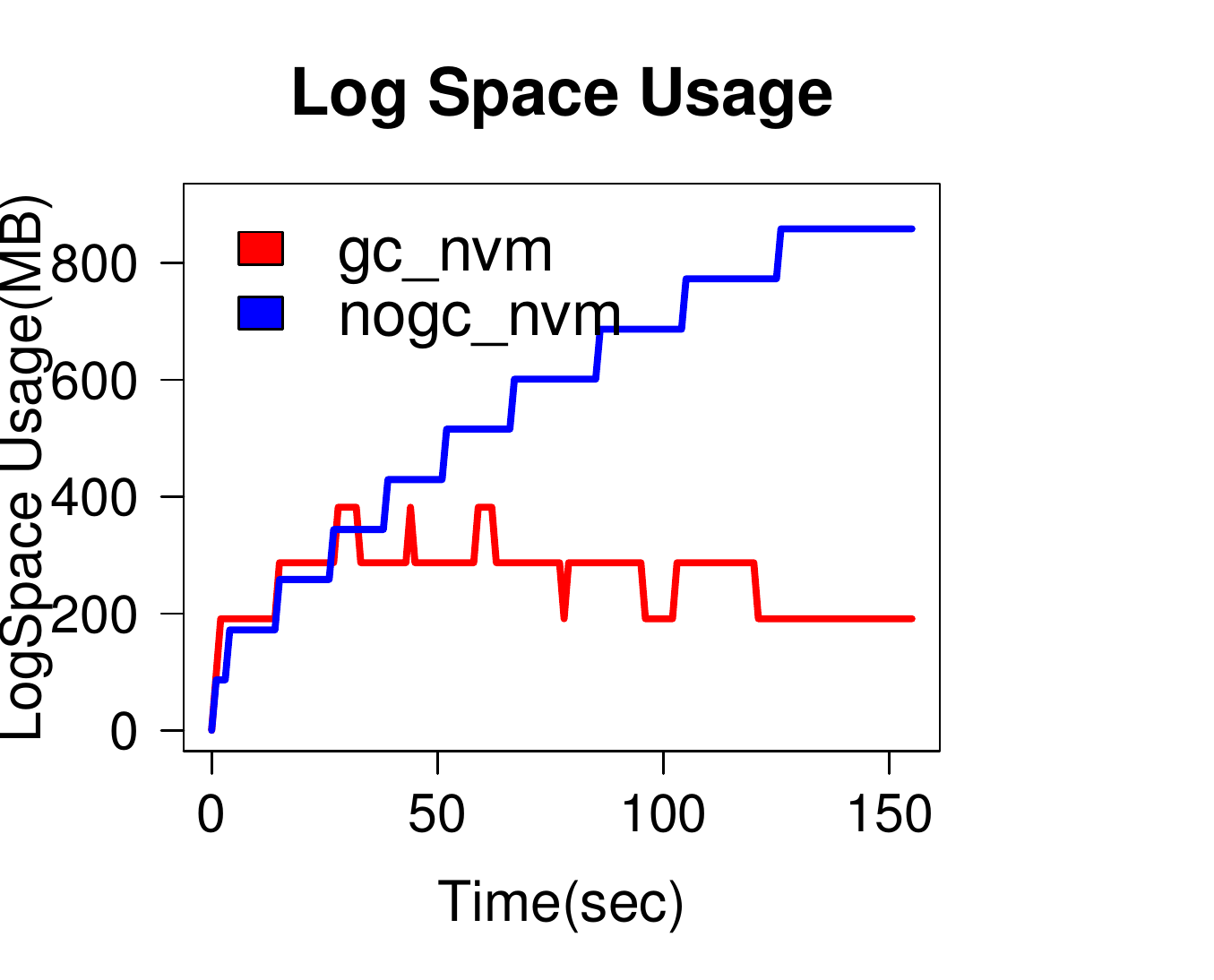}
						\caption{log space usage}
						\label{fig:pmemspace}
					\end{subfigure}
				\caption{The left plot shows the instant throughput with and without GC for both NVM and SSD. 
						The right plot shows the space usage with and without GC for NVM. Specifically, gc\_nvm means the system with 
						GC enabled and NVM based log; nogc\_nvm is the system with NVM based log but GC disabled; gc\_ssd
						is the system with GC enabled and SSD based log.}
			\end{center} 
	\vspace{-4mm} 
	\end{figure}
Two experiments are conducted to validate the effectiveness and efficiency of the NVM based log.
Both experiments are run with 36 servers and 95\% read transactions.
In Figure \ref{fig:pmemthroughput}, the instant throughput is plotted with and without
GC. The GC process doesn't affect normal transactions
when WALs are GC'ed every 10 seconds since
the reclamation of volatile states that affects normal transactions
is delayed until servers are in light loads. 
The system with SSD log shows 19k ops/sec on average, which is 30 times slower than that with NVM log. 
In Figure \ref{fig:pmemspace}, we measure the space over time with and without GC
to show the GC efficiency. The logs are NVM files of
100MB size so that the space usage changes in units of 100MB. 
The GC mechanism successfully keeps log space usage low since DTranx reclaims the 
transactions from WALs much faster than it writes them.
After 120 seconds, tests are complete and the log usage with GC
converges to 200 MB(one for GCLog and one for the current WAL).

\section{Related Work}
\label{sec:related}
DTranx is a highly concurrent and transactional KV store that integrates various 
techniques from concurrent programming, database, and NVM fields.

\textbf{Distributed Transaction.} Transaction research are heavily explored
in the database field and we investigated both classic and state-of-the-art
methods to guide the DTranx design. 
Spanner~\cite{spanner} was a globally consistent and efficient key value storage
system, which required atomic clocks to be installed on each server and its 
two-phase locking approach limited concurrency. Yang~\cite{chains} introduced
transaction chains to obtain both serializable transactions and low latency but
required that read and write items to be known as a priori, similar to Granola ~\cite{granola}. 
Calvin~\cite{calvin} designed a deterministic locking protocol to eliminate distributed commit protocols 
, but it enforced a global synchronization of transaction orders. SiloR~\cite{silor} used OCC but required
only exclusive locks on write items, while both shared and exclusive locks are requested in DTranx.
SiloR would require two successful rounds of operations (exclusive lock, followed by read);
in a distributed system, these two rounds would have significant latency due
to RPC calls, but SiloR was implemented on a single computer and did not use RPCs.
GMU~\cite{gmu} avoided
read only transaction aborts by guaranteeing the Extended Update
Serializability (EUS) isolation level, where read only transactions might observe
snapshots from different linearizations of update transactions. This might work for
some applications but as a fully ACID compliant KV store, DTranx enforces strong
isolation. Other approaches that added serializability to snapshot isolation such as
\cite{critique}, \cite{snapshot:serializable} required a central server for validation 
in distributed environment.

We explored various distributed transaction designs and chose the system that combined 
OCC and 2PC since it yielded great performance and guaranteed strong consistency without
special hardware such as atomic clocks. 

\textbf{Hardware-assisted Transactional System.}
With hardware advances like NVM, RDMA, software designs adopting these technologies
show tremendous performance growth. NVM devices, such as PCM(Phase Change Memory), 
3D XPoint, emerge and significantly reduce persistence overhead. Tianzheng et al.~\cite{Wang:2014:SLT:2732951.2732960}
designed a scalable log leveraging NVM to support distributed logging where they focused
on alleviating the contention bottleneck with passive group commit. According to our
experiments, distributed transactional systems based on NVM logs
yield such high throughput that the bottleneck resides in CPU processing.
Hence, we focus on building an efficient GC supported NVM log.
METRADB~\cite{metradb} was a middle layer that provided key value interfaces
for applications and hided the complexities of using NVML to facilitate application
development. However, we aim at providing an efficient garbage collection mechanism. 
On the other hand, HERD~\cite{herd} focused on building key
value services in memory using RDMA to reduce network round trips but lacked
fault recovery over server failures. In the future, we plan to explore RDMA for
lower latency RPCs.

\section{Conclusions}
We propose a transactional and scalable key value store
that utilizes non-volatile memory based log with an effective and
efficient garbage collection mechanism. To exploit the multi-core
machines, we adapt the SEDA architecture with lock free queues and
apply an optimal core binding strategy. Moreover, DTranx combines
OCC and 2PC to move the locks to the commit time and employ parallel
validation for better scalability. Experiments show that DTranx offers
higher throughput than the state-of-the-art system, Hyperdex, and DTranx displays
high scalability for various workloads.
\bibliography{reference}

\begin{thebibliography}{10}

\bibitem{pmem}
Intel {NVML}.
\newblock \url{https://pmem.io/}.

\bibitem{leveldb}
Leveldb -- a fast and lightweight key/value database library by google
  http://code.google.com/p/leveldb/.

\bibitem{strace}
Linux {strace}.
\newblock \url{https://strace.io/}.

\bibitem{sttmram}
D.~Apalkov, A.~Khvalkovskiy, S.~Watts, V.~Nikitin, X.~Tang, D.~Lottis, K.~Moon,
  X.~Luo, E.~Chen, A.~Ong, A.~Driskill-Smith, and M.~Krounbi.
\newblock Spin-transfer torque magnetic random access memory (stt-mram).
\newblock {\em J. Emerg. Technol. Comput. Syst.}, 9(2):13:1--13:35, May 2013.

\bibitem{concurrency_control}
P.~A. Bernstein and N.~Goodman.
\newblock Concurrency control in distributed database systems.
\newblock {\em ACM Comput. Surv.}, 13(2):185--221, June 1981.

\bibitem{snapshot:serializable}
M.~J. Cahill, U.~R\"{o}hm, and A.~D. Fekete.
\newblock Serializable isolation for snapshot databases.
\newblock In {\em SIGMOD}, pages 729--738, 2008.

\bibitem{BigTable}
F.~Chang, J.~Dean, S.~Ghemawat, W.~C. Hsieh, D.~A. Wallach, M.~Burrows,
  T.~Chandra, A.~Fikes, and R.~E. Gruber.
\newblock Bigtable: A distributed storage system for structured data.
\newblock In {\em Proceedings of the 7th USENIX Symposium on Operating Systems
  Design and Implementation - Volume 7}, OSDI '06, pages 15--15, Berkeley, CA,
  USA, 2006. USENIX Association.

\bibitem{ycsb}
B.~F. Cooper, A.~Silberstein, E.~Tam, R.~Ramakrishnan, and R.~Sears.
\newblock Benchmarking cloud serving systems with {YCSB}.
\newblock In {\em SOCC}, pages 143--154, 2010.

\bibitem{spanner}
J.~C. Corbett, J.~Dean, M.~Epstein, A.~Fikes, C.~Frost, J.~J. Furman,
  S.~Ghemawat, A.~Gubarev, C.~Heiser, P.~Hochschild, W.~Hsieh, S.~Kanthak,
  E.~Kogan, H.~Li, A.~Lloyd, S.~Melnik, D.~Mwaura, D.~Nagle, S.~Quinlan,
  R.~Rao, L.~Rolig, Y.~Saito, M.~Szymaniak, C.~Taylor, R.~Wang, and
  D.~Woodford.
\newblock Spanner: Google's globally distributed database.
\newblock {\em ACM Trans. Comput. Syst.}, 31(3):8:1--8:22, 2013.

\bibitem{granola}
J.~Cowling and B.~Liskov.
\newblock Granola: Low-overhead distributed transaction coordination.
\newblock In {\em USENIX ATC}, pages 223--235, 2012.

\bibitem{warp}
R.~Escriva and E.~G. Sirer.
\newblock The design and implementation of the warp transactional filesystem.
\newblock In {\em NSDI}, pages 469--483, 2016.

\bibitem{dbcomplete}
H.~Garcia-Molina.
\newblock {\em Database systems: the complete book}.
\newblock Pearson Education India, 2008.

\bibitem{GFS}
S.~Ghemawat, H.~Gobioff, and S.-T. Leung.
\newblock The google file system.
\newblock In {\em Proceedings of the Nineteenth ACM Symposium on Operating
  Systems Principles}, SOSP '03, pages 29--43, New York, NY, USA, 2003. ACM.

\bibitem{Harder:1984}
T.~H\"{a}rder.
\newblock Observations on optimistic concurrency control schemes.
\newblock {\em Inf. Syst.}, 9(2):111--120, Nov. 1984.

\bibitem{herd}
A.~Kalia, M.~Kaminsky, and D.~G. Andersen.
\newblock Using {RDMA} efficiently for key-value services.
\newblock {\em SIGCOMM Comput. Commun. Rev.}, 44(4):295--306, 2014.

\bibitem{cassandra}
A.~Lakshman and P.~Malik.
\newblock Cassandra: A decentralized structured storage system.
\newblock {\em SIGOPS Oper. Syst. Rev.}, 44(2):35--40, 2010.

\bibitem{metradb}
L.~Marmol, J.~Guerra, and M.~K. Aguilera.
\newblock Non-volatile memory through customized key-value stores.
\newblock In {\em USENIX HotStorage}, pages 101--105, 2016.

\bibitem{raft}
D.~Ongaro and J.~Ousterhout.
\newblock In search of an understandable consensus algorithm.
\newblock In {\em USENIX ATC}, pages 305--319, 2014.

\bibitem{gmu}
S.~Peluso, P.~Ruivo, P.~Romano, F.~Quaglia, and L.~Rodrigues.
\newblock When scalability meets consistency: Genuine multiversion
  update-serializable partial data replication.
\newblock In {\em ICDCS}, pages 455--465, 2012.

\bibitem{cloudlab}
R.~Ricci, E.~Eide, and {The CloudLab Team}.
\newblock Introducing {CloudLab}: Scientific infrastructure for advancing cloud
  architectures and applications.
\newblock {\em {USENIX} {;login:}}, 39(6), 2014.

\bibitem{docc2}
A.~Thomasian.
\newblock Distributed optimistic concurrency control methods for
  high-performance transaction processing.
\newblock {\em IEEE Transactions on Knowledge and Data Engineering},
  10(1):173--189, 1998.

\bibitem{calvin}
A.~Thomson, T.~Diamond, S.-C. Weng, K.~Ren, P.~Shao, and D.~J. Abadi.
\newblock Calvin: Fast distributed transactions for partitioned database
  systems.
\newblock In {\em SIGMOD}, pages 1--12, 2012.

\bibitem{Wang:2014:SLT:2732951.2732960}
T.~Wang and R.~Johnson.
\newblock Scalable logging through emerging non-volatile memory.
\newblock {\em PVLDB}, 7(10):865--876, 2014.

\bibitem{seda}
M.~Welsh, D.~Culler, and E.~Brewer.
\newblock {SEDA}: An architecture for well-conditioned, scalable internet
  services.
\newblock {\em SIGOPS Oper. Syst. Rev.}, 35(5):230--243, 2001.

\bibitem{critique}
M.~Yabandeh and D.~G\'{o}mez~Ferro.
\newblock A critique of snapshot isolation.
\newblock In {\em EuroSys}, pages 155--168, 2012.

\bibitem{chains}
Y.~Zhang, R.~Power, S.~Zhou, Y.~Sovran, M.~K. Aguilera, and J.~Li.
\newblock Transaction chains: Achieving serializability with low latency in
  geo-distributed storage systems.
\newblock In {\em SOSP}, pages 276--291, 2013.

\bibitem{silor}
W.~Zheng, S.~Tu, E.~Kohler, and B.~Liskov.
\newblock Fast databases with fast durability and recovery through multicore
  parallelism.
\newblock In {\em OSDI}, pages 465--477, 2014.

\end{thebibliography}
\bibliographystyle{abbrv} 

\end{document}